\newcommand{\beq}{\begin{equation}}
\newcommand{\eeq}{\end{equation}}
\newcommand{\bea}{\begin{eqnarray}}
\newcommand{\eea}{\end{eqnarray}}

\documentclass[twocolumn,prd,aps,amssymb,nofootinbib,epsf]{revtex4}
\usepackage{bm}
\usepackage{amsmath}
\usepackage{amsfonts}
\usepackage{graphicx}

\begin{document}

\title{Gravitational Radiation from Preheating with Many Fields}

\author{John T. Giblin, Jr${}^{1,2}$}
\author{Larry R. Price${}^3$}
\author{Xavier Siemens${}^3$}

\affiliation{${}^1$Department of Physics, Kenyon College, Gambier, OH 43022}
\affiliation{${}^2$Perimeter Institute for Theoretical Physics, 31 Caroline St. N, Waterloo, ON N2L 2Y5}
\affiliation{${}^3$Center for Gravitation and Cosmology, Department of
  Physics, University of Wisconsin--Milwaukee, P.O. Box 413, Milwaukee, WI  53201}
\date{\today}

\begin{abstract} 
  Parametric resonances provide a mechanism by which particles can be
  created just after inflation.  Thus far, attention has focused on a
  single or many inflaton fields coupled to a single scalar field.
  However, generically we expect the inflaton to couple to many other
  relativistic degrees of freedom present in the early universe.
  Using simulations in an expanding Friedmann-Lema\^itre-Robertson-Walker spacetime, in this paper we
  show how preheating is affected by the addition of multiple fields
  coupled to the inflaton. We focus our attention on gravitational
  wave production--an important potential observational signature of the
  preheating stage. We find that preheating and its gravitational wave
  signature is robust to the coupling of the inflaton to more matter fields.
\end{abstract}


\maketitle

\section{Introduction}

After inflation the universe is cold and far from thermal equilibrium.
Early ideas for how to reheat the universe after inflation involved
standard model particle production via oscillations of the inflaton
about the minimum of its
potential~\cite{Linde:1981mu,Albrecht:1982mp,Dolgov:1982th,Abbott:1982hn}.
The coupling between the inflaton and matter fields, however, can
lead to parametric amplification of the quantum fluctuations of the
matter fields. Thus, the reheating process can be preceded by a stage of
exponential particle production which has become known as
preheating~\cite{Traschen:1990sw,Kofman:1994rk,Kofman:1997yn}.

Inflation produces a stochastic background of gravitational waves
through the amplification of primordial quantum
fluctuations~\cite{Starobinsky:1979ty,Allen:1987bk}.  It turns out
that preheating may also lead to the production of a gravitational wave
background. This realization was first made by Khlebnikov and
Tkachev~\cite{Khlebnikov:1997di} in a quartic inflation model.  A
similar estimate by Garcia-Bellido somewhat
later~\cite{GarciaBellido:1998gm}, showed that in hybrid inflation the
frequency of the peak in the spectrum could be brought down to about
1~kHz, the frequency range where where current earth-based
interferometric gravitational-wave detectors operate.

This problem has received much attention
recently~\cite{Easther:2006gt,Easther:2006vd,Felder:2006cc,GarciaBellido:2007dg,Easther:2007vj,%
Dufaux:2007pt,GarciaBellido:2007af,Price:2008hq,Dufaux:2008dn}. Most of this work
uses {\sc LatticeEasy}~\cite{Felder:2000hq} to numerically evolve the inflaton and other
scalar fields in  Friedmann-Lema\^itre-Robertson-Walker (FLRW) spacetimes~\cite{Felder:2000hq}, but different
schemes to compute the metric perturbation and gravitational wave
spectrum. There is now general agreement on the gravitational wave
signal predicted by the preheating mechanism~\cite{Price:2008hq}.
All of this work used toy models with one inflaton field and one
(scalar) matter field. More recently there has been some progress in
our understanding of preheating with many inflaton 
fields~\cite{Battefeld:2008rd,Battefeld:2009xw,Barnaby:2010wd}, a possibility
suggested by string-theory-based inflation scenarios.

Generally, we expect the inflaton to couple to many of the 
degrees of freedom present in the early universe. Up until now, the effect of
these extra fields on the resonance and the gravitational wave
signature of preheating has not been explored.  Using simulations in an
expanding FLRW spacetime, in this paper we show
how preheating is affected by the addition of multiple fields coupled
to the inflaton, focusing our attention on gravitational wave production. 
We briefly discuss the possibility of determining the number of 
fields the inflaton is coupled to from the gravitational wave signature.

In Sec.~\ref{strategery} we review the dynamics of preheating and
outline our computational strategy to calculate the gravitational wave
spectrum. In Sec.~\ref{results} we discuss our results for quartic and
quadratic models of inflation, for the case when the inflaton is
coupled to between 1 and 32 matter fields. Throughout we compare
our results to the well-known case where the inflaton is coupled to
a single field. We conclude in Sec.~\ref{concl}.

\section{Model and Computational Method}
\label{strategery}

We will consider a simple extension of the usual preheating model \cite{Traschen:1990sw} in
which the inflaton, $\phi$, is coupled to a set of $\mathcal{N}$ scalar fields,
$\chi_\alpha$, according to the Lagrangian
\beq
{\cal L} =\sqrt{-g}\left( \frac{1}{2}\nabla_\mu\phi\nabla^\mu\phi + \frac{1}{2}\sum_{\alpha=1}^\mathcal{N}
\nabla_\mu\chi_\alpha\nabla^\mu \chi_\alpha  - V(\phi,\chi_\alpha) \right),\label{lagrangian}
\eeq
where $V(\phi,\chi_\alpha)$ contains both the inflationary potential, $V(\phi)$ and the interactions
between the inflaton and the scalars,
\begin{equation}
\label{potentialform}
V(\phi,\chi_\alpha) = V(\phi) + \sum_{\alpha=1}^\mathcal{N} \frac{1}{2} g^2 \phi^2\chi_\alpha^2.
\end{equation}
For simplicity we will assume that the coupling between the matter
fields and the inflaton, $g^2$, is the same for all fields. The fields
evolve in a FLRW spacetime, which takes the form
\begin{eqnarray}
\label{metric}
ds^2 &=& dt^2 - a^2(t) \left[dx^2+dy^2+dz^2\right]\\
&=& a^2(\tau)\left[d\tau^2 - dx^2-dy^2-dz^2\right],
\end{eqnarray}
in cosmological and conformal coordinates respectively.
In this paper we will consider both quadratic, $V(\phi) = m^2 \phi^2/2$,
and quartic, $V(\phi) = \lambda \phi^4/4$, inflationary models.

The fields obey the usual Klein-Gordon equations,
\begin{equation}
\ddot{\phi} + 3H \dot{\phi} - \frac{\nabla^2\phi}{a^2} + \frac{\partial V(\phi)}{\partial \phi} + \sum_{\alpha}^\mathcal{N} g^2 \phi \chi_{\alpha}^2 = 0,
\end{equation}
and
\begin{equation}
\ddot{\chi}_{\alpha} + 3H \dot{\chi}_{\alpha} - \frac{\nabla^2\chi_{\alpha}}{a^2} + g^2 \phi^2 \chi_{\alpha} = 0. \label{chieom}
\end{equation}
To show that we can achieve parametric resonance in this model, it's easiest to recast  Eq.~(\ref{chieom}) in terms of the mode equations of $\chi_{\alpha}$,
\begin{equation}
\label{mode}
\ddot{\tilde{\chi}}_{\alpha} + 3H \dot{\tilde{\chi}}_{\alpha} +\left(\frac{k^2}{a^2} + g^2 \phi^2 \right)\tilde{\chi}_{\alpha} = 0,
\end{equation}
where we use the Fourier convention,
\begin{equation}
\tilde{f}(\vec{k},t) = \int d^3x \, f(\vec{x},t) e^{2\pi i \vec{k} \cdot \vec{x}}.
\end{equation}

Before we extend our analysis to many fields, we will briefly review
the phenomenon of parametric resonance.  In both quadratic and quartic
inflation, the end of inflation is marked by coherent oscillations of
the $\phi$ field.  At the end of quadratic inflation, temporarily
ignoring the expansion of the universe, the inflation will oscillate
sinusoidally,
\begin{equation}  
\phi(\vec{x},t) = \Phi \sin (mt).
\end{equation}
In the case of a single coupled scalar, $\chi$, the mode equations
Eq.~(\ref{mode}) can be cast into a Mathieu equation \cite{bateman},
\begin{equation}
\label{mathieu}
\tilde{\chi}^{\prime\prime} + \left(A_k - 2q\cos(2z)\right)\tilde{\chi} = 0
\end{equation}
using the coordinate transformation,
\begin{equation}
\label{coordtransform}
q=\frac{g^2\Phi^2}{4m^2}, \,A_k = \frac{k^2}{m^2}+\frac{g^2\Phi^2}{2m^2},\,z=mt,
\end{equation}
where primes denote differentiation with respect to $z$.  Solutions to
Eq.~(\ref{mathieu}) can be characterized as either stable (oscillatory) or
unstable (exponential) depending on the choice of
parameters\footnote{See, for example, \cite{Easther:2007vj} for
  stability diagrams.}, $A_k$ and $q$.  Preheating, in the form of
parametric resonance, occurs when particular modes of $\chi$ are
excited, producing a universe whose content is out of thermal
equilibrium \cite{Kofman:1994rk,Kofman:1997yn}.  In practice, the inclusion of the expansion of
the universe complicates the coordinate transformation
Eq.~(\ref{coordtransform}) by introducing time-dependent Mathieu
parameters, $A_k(t)$ and $q(t)$ \cite{Kofman:1994rk,Kofman:1997yn}. 

In the case of quartic inflation, the time dependence of $\phi$ is not
purely sinusoidal; rather, the field solution can be written in terms of an elliptic
cosine
\begin{equation}
\phi \propto \mathrm{cn} \left(x-x_0,\frac{1}{\sqrt{2}}\right),
\end{equation}
where $x \propto \sqrt{t}$ \cite{Greene:1997fu}.  Substituting this
into the single-field version of Eq.~(\ref{mode}) yields a Lam\'e
equation.  The dynamics of this can be understood as an approximate
Mathieu equation since the elliptical cosine can be decomposed into
sinusoidal parts, each of which have an associated Mathieu equation
\cite{Greene:1997fu}.

This process is non-thermal and generates large anisotropies in the
stress-energy tensor. Khlebnikov and Tkachev predicted
that this anisotropic stress-energy should be an efficient generator
of stochastic gravitational waves \cite{Khlebnikov:1997di}.  The
problem remained largely ignored for several years until Easther
and Lim confirmed the effect~\cite{Easther:2006gt}, in work that sparked a resurgence of
interest and activity in the field~\cite{GarciaBellido:2007dg,Easther:2007vj,
Dufaux:2007pt,GarciaBellido:2007af,Price:2008hq,Dufaux:2008dn}.

In principle, our understanding of parametric resonance can be
trivially extended to the case of many scalar fields, $\chi_{\alpha}$.  Since
we are taking a single degree of freedom for the inflaton, the
behavior of $\phi(t)$ in our model Eq.~(\ref{lagrangian}), at least initially, should be
identical to the single-scalar case.  The only difference being that
there will be $\mathcal{N}$ identical Mathieu/Lam\'e equations,
parametrically sourcing inhomogeneities in each of the $\chi_{\alpha}$.

We do not expect trivial behavior in the dynamics of multi-scalar {\sl
  preheating}.  In the case of a single scalar,
parametric resonance continues until the backreaction becomes
important and the inhomogeneities in $\chi$ cause the inflaton to
loose coherence via the interaction potential $V(\phi, \chi)$.  In the case of
many scalars, we expect this effect to be enhanced.

Although the process is highly nonlinear, qualitatively we
expect the introduction of many scalar fields  $\chi_{\alpha}$ to have an effect on the
efficiency of preheating either by shortening the process or limiting
the amplification of the modes of any particular $\chi_{\alpha}$.  The open
questions are: (1) how different are the nonlinear dynamics of
preheating in the case of many-scalars, and (2) what effect will this
have on the generation of gravitational radiation.

To address these questions, we have performed a series of numerical 
simulations with $\mathcal{N}=1,2,4,8,16,$ and $32$ matter fields  
$\chi_{\alpha}$.  We use the publicly available early universe field evolution package
{\sc LatticeEasy} to perform our simulations. We
couple {\sc LatticeEasy} to a code that evolves the metric perturbation
using the methods of~\cite{Easther:2006vd,Easther:2007vj}.

\begin{figure}[htbp] 
   \centering
   \includegraphics[width=3in]{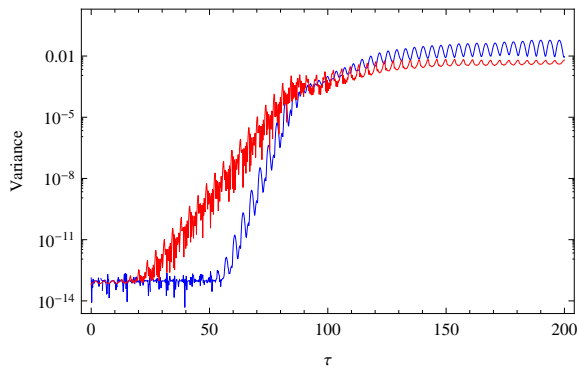} 
   \caption{The field variances, $\phi$ (blue) and $\chi$ (red), as a function of conformal time, $\tau$, for the $\mathcal{N}=1$ case. In this simulation the inflaton potential is quartic with $\lambda=10^{-14}$ and $g^2/\lambda = 120$.}
   \label{fig:variances2}
\end{figure}

Tensor perturbations to the metric, $h_{ij}$, can be written in the synchronous gauge 
\begin{equation}
\label{sync}
ds^2 = dt^2 - a^2(t)\left[\delta_{ij} + h_{ij}\right]dx^idx^j.
\end{equation}
For this choice of gauge perturbations are purely spatial and obey the transverse-traceless conditions
\begin{equation}
h_i^i = 0\,\,\,{\rm and} \,\,\,\,h_{ij,j} =0.
\end{equation}
The metric perturbations obey sourced Klein-Gordon equations
\begin{equation}
\ddot{h}_{ij} + 3H \dot{h}_{ij} + \frac{1}{a^2} \nabla^2 h_{ij} = 16 \pi S^{TT}_{ij}
\label{equationofmotion}
\end{equation}
where the source term, $S^{TT}_{ij}$ is the transverse-traceless part of the anisotropic stress tensor,
\begin{equation}
\label{aniso}
S_{ij} = T_{ij} - \frac{\eta_{ij}}{3}T.
\end{equation}
We work in natural units where the speed of light and Newton's gravitational constant are set to unity, $c=G=1$.

After we specify our model Eq.~(\ref{lagrangian}), {\sc LatticeEasy}
evolves the fields and the scale factor.  From this, we calculate the
momentum-space anisotropic stress tensor, Eq.~(\ref{aniso}), which we use
to source the evolution of the metric perturbation $h_{ij}$.

We compute the present-day power spectrum of stochastic gravitational radiation coming from preheating as follows. The stress energy tensor associated with the metric perturbations is \cite{Misner:1974qy}, 
\begin{equation}
T^{\rm gw}_{\mu\nu} = \frac{1}{32\pi} \left<h_{ij,\mu}h^{ij}_{\,\,\,,\nu}\right>,
\end{equation}
where the brackets denote a spatial average.  
The energy density in gravitational waves is given by 
\begin{equation} 
\rho_{\rm{gw}} = \frac{t^{\mu}t^{\nu}}{32\pi} \left<h_{ij,\mu}h^{ij}_{,\nu}\right> = \frac{1}{32 \pi} \sum_{i,j}
 \left<\dot{h}^2_{ij}\right>, \label{gwdensity}
\end{equation}
where $t^{\mu} = (1,0,0,0)$ in the background metric defined in Eq.~(\ref{sync}) and the overdot denotes
a time derivative.
Using Parseval's theorem (see \cite{Easther:2007vj}) we can compute the power spectrum at the end of the simulation,
\begin{equation}
\rho_{\rm gw} =\frac{1}{32\pi} \frac{1}{V}\sum_{i,j}\int d^3\mathbf{k}\,\, 
\Bigl|\dot{h}_{ij}(t,\mathbf{k})\Bigr|^2
\label{omega0}
\end{equation}
where $V$ is the comoving volume over which the
spatial average is being performed. We can then write
\begin{equation}
\frac{d\rho_{\rm gw}}{d\ln k} = 
\frac{k^3}{32\pi} \frac{1}{V} \sum_{i,j} \int d\Omega\,
\Bigl| \dot{h}_{ij}^{\rm TT}(\eta,\mathbf{k}) \Bigr|^2,
\label{omega}
\end{equation}
which is related to the present-day power spectrum via~\cite{Easther:2007vj,Price:2008hq},
\begin{equation}
\Omega_{\rm gw,0}h^2 = \Omega_{\rm rad,0}h^2 \Biggl(\frac{g_0}{g_e}\Biggr)^{1/3}
\frac{1}{\rho_{\rm tot, e}}\frac{d\rho_{\rm gw,e}}{d\ln k}
\label{omega1}
\end{equation}
where the 0 and e subscripts denote quantities today and the end of
our simulations respectively, $h$ is a dimensionless factor that
absorbs the uncertainty in the present value of the Hubble parameter,
$\Omega_{\rm rad,0}$ is the current fraction of the energy density in
the form of radiation, and $\rho_{\rm tot, e}$ is the total energy
density at the end of our simulations. The ratio, $g_0/g_{\rm e}$, is
the number of degrees of freedom today to the number of degrees of
freedom at matter/radiation equality.  We approximate $g_0/g_{\rm e}=1/100$.

Our simulations run on a 3-dimensional lattice with $128$ points along each direction.

\section{Results}
\label{results}

In both massive and massless models, the period immediately following inflation is
characterized by the coherent oscillation of the inflaton, $\phi$,
about its minimum.  Since this is the dominant mode, during this period the inflaton acts as a
homogeneous field $\Phi(t)$.  As such, the $\chi_\alpha$ fields each obey
identical equations of motion,
\begin{equation}
\ddot{\chi}_{\alpha} + 3 H \dot{\chi}_{\alpha} - \frac{\nabla^2\chi_{\alpha}}{a^2} + g^2\Phi^2(t)\chi_\alpha = 0.
\label{homogeneous}
\end{equation}
We expect the $\chi_\alpha$ fields to undergo resonance during this
stage, each field acting identically.  

Our expectations for the dependence on $\mathcal{N}$ for the gravitational wave spectrum during this period can be sketched out with the following simple argument.  
The metric perturbation is given by
\begin{equation}
h_{ij}(t,\mathbf{k}) \propto  \int\, dt' G(t,t')T_{ij}^{\rm TT}(t',\mathbf{k}),
\end{equation}
where $G(t,t')$ is the Green's function and $t$ is the appropriate time coordinate, either conformal or co-moving, and
the overdot represents a time derivative.  The gravitational wave spectrum is
\begin{eqnarray}
\Omega_{\rm gw} &\propto& \sum_{i,j}\frac{d}{d\ln k}\left\langle\dot{h}_{ij}^2\right\rangle \nonumber \\
&\propto& \sum_{i,j}\int\, d\Omega dt' dt'' \dot{G}(t,t')\dot{G}(t,t'') \nonumber \\
&\phantom{\propto}&\phantom{\sum_{i,j}\int\,} \times T_{ij}^{\rm TT}(t',\mathbf{k})T_{ij}^{*\,{\rm TT}}(t'',\mathbf{k}).
\end{eqnarray}
For each direction $\Omega$ in the integral above, we can perform a rotation to a frame where the gravitational wave
is traveling along the $z$-direction.  In this case the square of the stress-energy
tensor has a particularly simple form
\begin{equation}
\label{bigsum}
\sum_{i,j}T_{ij}^{\rm TT}T_{ij}^{*\,{\rm TT}} = \left( \left|(T_{xx} - T_{yy})/2\right |\right)^2 + \left( \left| T_{xy}\right |\right)^2.
\end{equation}
Every component of the stress energy tensor is a sum of contributions from each field,
\beq
T_{ij} = \sum_{\alpha=1}^\mathcal{N} \nabla_i \chi_\alpha \nabla_j\chi_\alpha - g_{ij}\left[\nabla_\mu \chi_\alpha \nabla^\mu \chi_\alpha+ V(\phi,\chi_\alpha)\right].
\eeq
Since the field values are uncorrelated we expect the sum to grow like $\sqrt{\mathcal{N}}$, and hence Eq.~(\ref{bigsum}) and the gravitational wave spectrum should scale like $\mathcal{N}$.

After this initial period, the evolution is highly nonlinear and we resort to numerical simulations to study it.  In the following two subsections we investigate the cases of quartic and quadratic inflation.

\subsection{Quartic inflation}

We start by examining massless inflation.  The inflationary potential is
\beq
V(\phi) = \frac{1}{4} \lambda \phi^4, 
\eeq
where we take the value of the self-coupling, $\lambda = 10^{-14}$, and $g/\lambda = 120$.  These particular 
values are chosen to make comparisons with \cite{Easther:2007vj,Dufaux:2007pt,Price:2008hq} simple and also 
to satisfy the COBE normalization \cite{Salopek:1992zg}.

\begin{figure}[htbp] 
   \centering
   \includegraphics[width=3in]{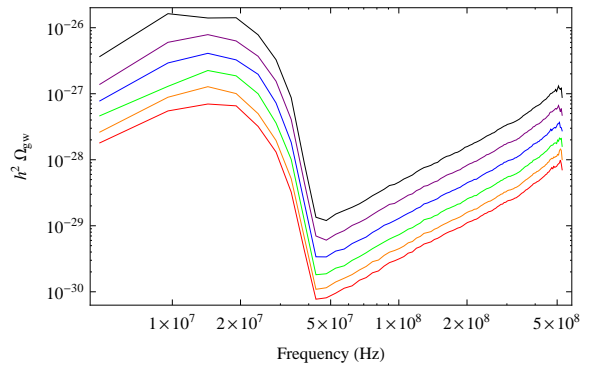} 
   \caption{The present day gravitational wave spectrum produced during the stage when the $\phi$ field is coherently oscillating.  The color coding is: $\mathcal{N}=1$ (red), $\mathcal{N}=2$ (orange), $\mathcal{N}=4$ (green), $\mathcal{N}=8$ (blue), $\mathcal{N}=16$ (purple), $\mathcal{N}=32$ (black).  In these simulations the inflaton potential is quartic with $\lambda=10^{-14}$ and $g^2/\lambda = 120$}
   \label{spectregone}
\end{figure}

The dynamics observed in our simulations can be described in three stages.  During the first stage the inflaton is 
oscillating coherently at the bottom of its potential, inducing parametric resonance in each matter field 
independently.  The gravitational wave spectrum at this point in the simulations is shown in 
Fig.~\ref{spectregone}.    
At both high and low frequencies the expected scaling of the gravitational wave spectrum proportional to $\mathcal{N}$ is observed.  Overall the gravitational wave production at 
this stage is not substantial. The peaks of the spectra will rise roughly 15 orders of magnitude before the end 
of the preheating process.  

\begin{figure}[htbp] 
   \centering
   \includegraphics[width=3in]{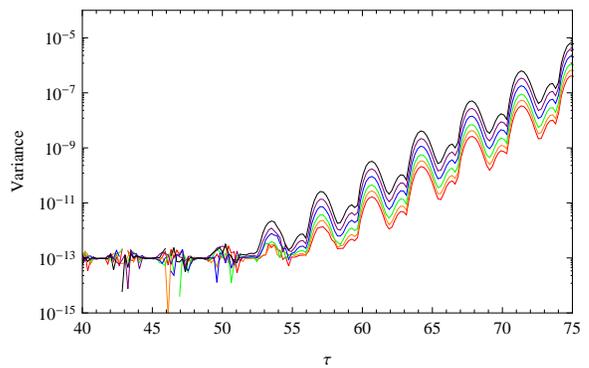} 
   \caption{The variances of $\phi$ for each simulation as a function of conformal time, $\tau$, during the stage of coherent oscillations of $\phi$.  The color coding is: $\mathcal{N}=1$ (red), $\mathcal{N}=2$ (orange), $\mathcal{N}=4$ (green), $\mathcal{N}=8$ (blue), $\mathcal{N}=16$ (purple), $\mathcal{N}=32$ (black).  In these simulations the inflaton potential is quartic with $\lambda=10^{-14}$ and $g^2/\lambda = 120$.}
   \label{phistage2}
\end{figure}

\begin{figure}[htbp] 
   \centering
   \includegraphics[width=3in]{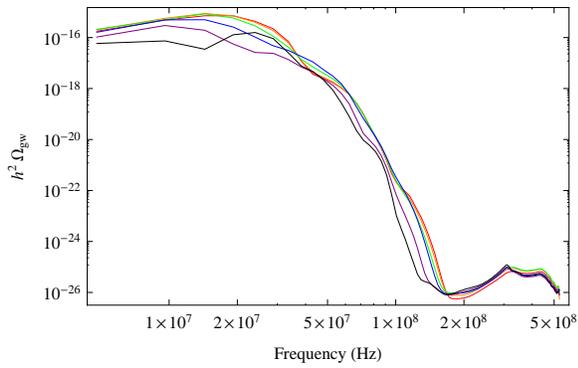} 
   \caption{The present day gravitational wave spectrum produced during the stage when the inhomogeneities of $\phi$ are growing, and the $\chi$ fields are still resonating.  The color coding is: $\mathcal{N}=1$ (red), $\mathcal{N}=2$ (orange), $\mathcal{N}=4$ (green), $\mathcal{N}=8$ (blue), $\mathcal{N}=16$ (purple).  In these simulations the inflaton potential is quartic with $\lambda=10^{-14}$ and $g^2/\lambda = 120$.}
   \label{spectregtwo}
\end{figure}

\begin{figure}[htbp] 
   \centering
   \includegraphics[width=3in]{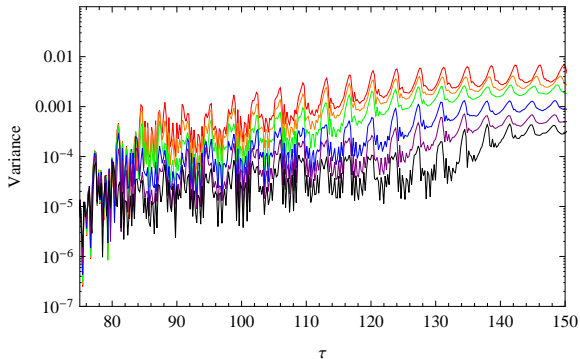} 
   \caption{The variances of one $\chi_{\alpha}$ for each simulation as a function of conformal time, $\tau$, during the stage when preheating becomes inefficient.  The color coding is: $\mathcal{N}=1$ (red), $\mathcal{N}=2$ (orange), $\mathcal{N}=4$ (green), $\mathcal{N}=8$ (blue), $\mathcal{N}=16$ (purple).}
   \label{chistage3}
\end{figure}

The next stage of the evolution is highly nonlinear. It is marked by the amplification of perturbations of the inflaton driven by the matter fields.  
The variances of the inflaton fields from each of the simulations are shown in Fig.~\ref{phistage2}.  
The 
simulations with more matter fields create more backreaction on the inflaton, causing the largest fluctuations
in the inflaton in the $\mathcal{N}=32$ simulation, followed by the $\mathcal{N}=16$ simulation, and so on.  Gravitational wave production at this stage is 
significant, as finally summarized in Fig.~\ref{spectregtwo}, 
but the peaks of the spectra are still approximately five orders of 
magnitude away from their final values.  At the same same time,
the increased backreation decreases the efficiency of the resonance and the result is that the inflaton 
resonance ends sooner in the simulations with more matter fields.  The effect on the matter fields is
shown in Fig.~\ref{chistage3}, where the effect of the different $\mathcal{N}$ becomes apparent in the evolution of the matter fields.

The nonlinearity of this stage makes it difficult to produce an analytic argument to understand the 
details of the $\mathcal{N}$-dependence.  In the final stage, the fields settle and, in this model of inflation, thermalize.  
  
\begin{figure}[htbp] 
   \centering
   \includegraphics[width=3in]{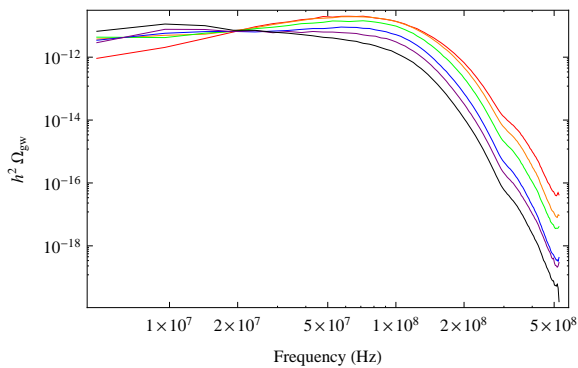} 
   \caption{The present day gravitational wave spectrum from preheating.  The color coding is: $\mathcal{N}=1$ (red), $\mathcal{N}=2$ (orange), $\mathcal{N}=4$ (green), $\mathcal{N}=8$ (blue), $\mathcal{N}=16$ (purple), $\mathcal{N}=32$ (black).  In these simulations the inflaton potential is quartic with $\lambda=10^{-14}$ and $g^2/\lambda = 120$.}
   \label{spectregthree}
\end{figure}

The final gravitational wave spectra are shown in Fig.~\ref{spectregthree}.  
Note that during this 
stage the higher frequency modes are excited more in the simulations with fewer fields and this leads to a crossing over over of the spectra.

\subsection{Quadratic inflation}

\begin{figure}[htbp] 
   \centering
   \includegraphics[width=3in]{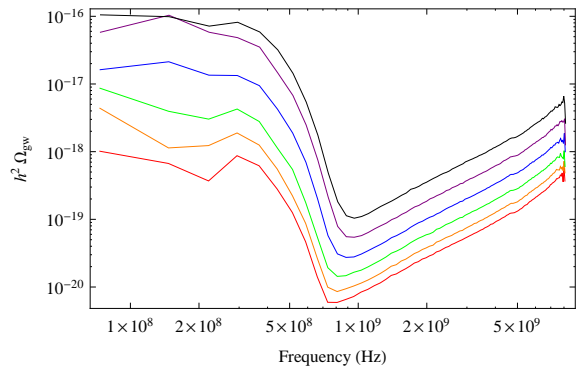} 
   \caption{The present day gravitational wave spectrum produced during the stage when the $\phi$ field is coherently oscillating.  The color coding is: $\mathcal{N}=1$ (red), $\mathcal{N}=2$ (orange), $\mathcal{N}=4$ (green), $\mathcal{N}=8$ (blue), $\mathcal{N}=16$ (purple), $\mathcal{N}=32$ (black).  In these simulations the inflation potential is quadratic with $m= 10^{-6}m_{\rm pl}$ and $g^2/m^2 = 2.5\times 10^5$.}
   \label{spectmeone}
\end{figure}

Lastly we investigate quadratic inflation.  The potential is given by
\beq
V(\phi) = \frac{1}{2} m^2 \phi^2,
\eeq
where $m= 10^{-6}m_{\rm pl}$ and $g^2/m^2 = 2.5\times 10^5$.  The stages of preheating are identical to those discussed in the previous subsection.  At the end of inflation the inflaton, $\phi$, is coherently oscillating and the modes of the matter fields, $\chi_\alpha$, are subject to (approximate) Mathieu equations.  During this stage the $\chi_\alpha$ are parametrically amplified by the coherent oscillations of the inflaton, and we expect the gravitational radiation produced to be proportional to the number of fields.  This can be seen in Fig.~\ref{spectmeone}.

\begin{figure}[htbp] 
   \centering
   \includegraphics[width=3in]{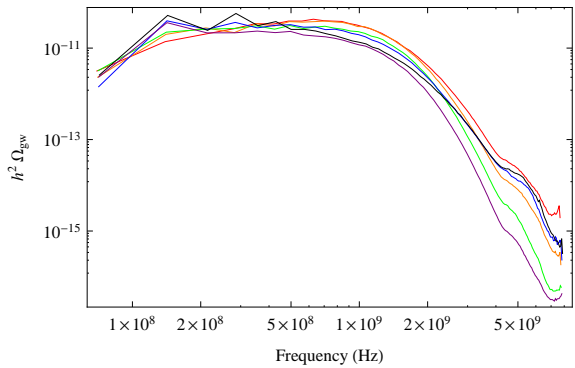} 
   \caption{The present day gravitational wave spectrum from preheating.  The color coding is: $\mathcal{N}=1$ (red), $\mathcal{N}=2$ (orange), $\mathcal{N}=4$ (green), $\mathcal{N}=8$ (blue), $\mathcal{N}=16$ (purple), $\mathcal{N}=32$ (black).  In these simulations the inflation potential is quadratic with $m= 10^{-6}m_{\rm pl}$ and $g^2/m^2 = 2.5\times 10^5$.}
   \label{spectmetwo}
\end{figure}

As preheating continues, the back-reaction on $\phi$ becomes important and preheating becomes inefficient.  The $\mathcal{N}$-dependent amplification of gravitational radiation is counteracted by this back-reaction and models with more fields cease to be more efficient.  By the time the source term of Eq.~(\ref{equationofmotion}) vanishes the spectrum of gravitational radiation is virtually independent of $\mathcal{N}$, see Fig.~\ref{spectmetwo}.  Namely, the peak frequency and amplitude remain roughly unchanged.

\subsection{$\mathcal{N}$ and g}

The form of the potential Eq.~(\ref{potentialform}) allows for the field redefinition,
\begin{equation}
\xi^2 = \sum_{\alpha} \chi_{\alpha}^2,
\end{equation}
along with a set of angles, $\theta_{\alpha}$, that cover the ($\mathcal{N}-1$)-sphere.  Using this set of fields, the inflation only couples to the norm of the field, $\xi$,
\begin{equation}
V(\phi,\xi) = V(\phi) + \frac{1}{2}g^2\mathcal{N}\phi^2\xi^2.
\end{equation}
This can lead one to believe that the addition of scalar degrees of freedom is equivalent to  changing the coupling in the case of a single field.  The analogy is not exact, though, since the kinetic term must also be modified.  Consider, for example, the case of $\mathcal{N}=2$.  In this case we only have one angle $\theta$ and the kinetic term becomes, 
\begin{equation}
\mathcal{T}  =\frac{1}{2}\nabla_\mu\phi\nabla^\mu\phi +\frac{1}{2}\nabla_\mu\xi\nabla^\mu\xi +\frac{1}{2}\xi^2 \nabla_\mu\theta \nabla^\mu \theta.
\end{equation}
In other words, if we wish to draw an analogy with single-scalar dynamics, we need to absorb the angular kinetic term into an effective potential,
\beq
V_{\rm eff} (\phi,\xi) = V(\phi) + \frac{1}{2}\left[g^2\mathcal{N}\phi^2 - \nabla_\mu\theta \nabla^\mu \theta\right]\xi^2.
\eeq
Using this approach, we can see that $\xi$ has a time-dependent mass term, and will generically {\sl not} vanish when $\phi$ vanishes.  We can only see one last way to force our analogy to be exact; namely to set $\theta=0$.  Outside of this limiting case, $\theta$ is a a dynamical degree of freedom and influences $\xi$.

\section{Conclusions}
\label{concl}

In the early universe, we expect the inflaton to be coupled to many of the fields present at that time. We have investigated the effect of these fields on the dynamics of preheating, and explored changes to the gravitational wave signature compared to the case with just one field--the only case investigated in the literature so far.

Our simulations show that parametric resonance is robust to the presence of many scalar fields coupled to the inflaton. In addition, we find significant gravitational wave production regardless of the number of fields.

In the case of many scalars we see two subtle, $\mathcal{N}$-dependent effects.  In both quartic and quadratic inflation models we observe an enhancement in the low-frequency content of the gravitational wave spectrum. This enhancement is caused by the presence of multiple $\chi_{\alpha}$ fields, and is sourced when the $\phi$ field is homogeneous and oscillating coherently. In this early phase the low frequency enhancement to the gravitational wave spectrum is proportional to  $\mathcal{N}$.
This is the stage when parametric resonance is most efficient.  At later times the presence of extra matter fields induces significant back-reaction on the inflaton and causes parametric resonance to lose efficiency relative to the $\mathcal{N}=1$ case. In the quartic inflation model this effect suppresses the high frequency content of the gravitational wave spectrum.  The combination of the low frequency enhancement and high frequency suppression results in a slight flattening of the spectrum when the number of matter fields is large. Conversely, the effect of the many matter fields on the gravitational wave spectrum in quadratic inflation is significantly less pronounced.

In \cite{Easther:2006vd} the authors claim that the detection of a preheating signal is complementary to the detection of 
primordial gravitational radiation from inflation, and that the energy scale of inflation can also be determined from the gravitational wave signature of preheating. The results of the present work further strengthen the claim that the gravitational wave signature is determined largely by the energy scale of inflation and only weakly depends on the number of matter fields that are coupled to the inflaton.

\section{Acknowledgments}

We would like to thank Jaume Garriga for suggesting this problem. We
are also grateful to Jolien Creighton, Richard Easther, Mustafa Amin, and Eugene Lim for
extremely useful discussions.  JTG would like to thank the
UW--Milwaukee for its generous hospitality while much of this work was
completed.  Research at the Perimeter Institute for Theoretical Physics is supported by the Government of Canada through Industry Canada and by the Province of Ontario through the Ministry of Research \& Innovation.  LP is supported by NSF Grant Nos. PHY-0503366,
PHY-0758155, and the Research Growth Initiative at the University of
Wisconsin- Milwaukee.  XS is supported in part by NSF Grant No.
PHY-0758155.

\bibliography{refs}

\end{document}